\documentclass[pra,aps,amssymb,twocolumn,noshowpacs]{revtex4-1}
\usepackage{color}
\usepackage{graphicx}
\usepackage{hyperref}
\usepackage{amsmath}
\usepackage{latexsym}
\usepackage{amssymb}
\usepackage{mathrsfs}
\usepackage{mathtools}
\usepackage{layout}
\usepackage{verbatim}
\usepackage{bm}
\usepackage{amsfonts,epsfig}

\usepackage{tikz}
\newcommand\circled[1]{\tikz[baseline=(char.base)]{
            \node[shape=circle,draw,inner sep=0.5pt] (char) {#1};}}
    
\begin{document}
\title{Accurate Quantum Logic Gates by Spin Echo in Rydberg Atoms}

\date{\today}
\author{Xiao-Feng Shi}
\affiliation{School of Physics and Optoelectronic Engineering, Xidian University, Xi'an 710071, China}

\begin{abstract}
Scalable quantum computing is based on realizable accurate quantum gates. For neutral atoms, it is an outstanding challenge to design a high-fidelity two-qubit entangling gate without resorting to difficult techniques like shaping laser pulses or cooling atoms to motional ground states. By using spin echo to suppress the blockade error, we propose an easily realizable controlled-phase Rydberg quantum gate of high intrinsic fidelity. In the context of spin echo, we show that the fundamental blockade error of the traditional Rydberg gate, on the order of $\epsilon\sim10^{-3}$, actually results from two `clockwise' rotations of Rabi frequencies $\bar\Omega_\pm=V\pm \sqrt{V^2+\Omega^2}$. In our  `echo' sequence, such an error can be suppressed to the order of $\epsilon^2$ by adding two `anticlockwise' rotations with frequencies $-\bar\Omega_\pm$. With the blockade error effectively removed, the error caused by Rydberg state decay becomes the final fundamental limit to the gate accuracy, which in principle, can be reduced beyond the level of $10^{-5}$. Furthermore, due to the small population $\epsilon$ involved in the `echo' process, the spin-echo gate is robust against the variation of Rydberg blockade caused by the drift of the qubits, so that it can still be much more accurate than that of a traditional Rydberg gate even for qubits cooled only to the sub-mK regime.

\end{abstract}
\maketitle

\section{introduction}
Quantum coherence and interference lie at the heart of quantum theory and the path integral formulation of quantum mechanics. A typical example of quantum coherence is the spin echo in a many-particle system~\cite{PhysRev.80.580,PhysRevLett.13.567,Slichter1992}, which is specifically important for quantum information processing in solid-state systems where many-body noise can be partly suppressed by spin echo~\cite{Liu2010}. Analogous spin echo, however, was rarely applied for neutral atoms~\cite{Zeiher2016,Zeiher2017}. In this work, we study an alternative spin echo in neutral Rydberg atoms and apply it to design an accurate entangling quantum gate.

Although neutral atom network is a promising platform for achieving large-scale quantum computing~\cite{Saffman2010,Saffman2016,Weiss2017}, it is not easy to design high-fidelity two-qubit Rydberg logic gates~\cite{Shi2017}, for which pulse shaping was usually assumed in theoretical proposals~\cite{Goerz2014,Theis2016,Petrosyan2017}. In contrast, our spin-echo method only depends on the Rydberg blockade mechanism, which was first proposed in~\cite{PhysRevLett.85.2208} and extensively studied in experiments~\cite{Wilk2010,Isenhower2010,Zhang2010,Maller2015,Jau2015,Zeng2017}.

The essence of our theory can be understood by considering a generic quantum system with two basis states $\{|\uparrow\rangle, |\downarrow\rangle\}$ subject to a time evolution under the control of the following Hamiltonian in a rotating frame
\begin{eqnarray}
 \hat{H}&=&
     V|\uparrow\rangle \langle\uparrow| +\Omega( |\uparrow\rangle \langle\downarrow|+\text{H.c.} )/2.\label{H00}
\end{eqnarray}
Diagonalization of the above Hamiltonian shows that starting from the state $|\psi(0)\rangle=|\downarrow\rangle$, the population in $|\downarrow\rangle$ will generally not return to 1 because the time evolution of $ |\psi(t)\rangle$ contains two quantum oscillations with irrational Rabi frequencies $\bar\Omega_\pm=V\pm \sqrt{V^2+\Omega^2}$~\cite{Shi2017}. In analogy to spin echo, $\bar\Omega_\pm$ represent two mismatched precession rates that result in an inhomogeneous broadening witnessed by $|\langle \downarrow|\psi(t)\rangle |^2<1$ when $t>0$. However, if the sign of every matrix element in Eq.~(\ref{H00}) is suddenly flipped after a duration of $t_0$, $\hat{H}\rightarrow -\hat{H}$, the two Rabi frequencies $\bar\Omega_\pm$ will be flipped to $-\bar\Omega_\pm$. Then, the state evolution in the following period of $t_0$ is exactly time reversed, leading the system back to its initial state at a later time $2t_0$, i.e., $|\psi(2t_0)\rangle= |\psi(0)\rangle$. This picture of spin echo can be extended to an N-level quantum system, where a given initial state can also experience time-reversed dynamics if the Hamiltonians are $\hat{H}_N$ and $\hat{H}_N'=- \hat{H}_N$ during the periods $t\in(0,t_0]$ and $t\in(t_0,2t_0]$, respectively. More generally, if the `anticlockwise' rotation is instead governed by a Hamiltonian $\hat{H}_N'=-|\varkappa| \hat{H}_N$, an echo can occur at the time $2t_0/|\varkappa|$, i.e., $|\psi(2t_0/|\varkappa|)\rangle= |\psi(0)\rangle$. This exotic dynamics can be simulated with Rydberg dressed neutral atoms, where the atom-light interaction of each atom is detuned by a large value of $\Delta$~\cite{Zeiher2016,Zeiher2017}: (i) For a duration $t_0$, optically dress a Rydberg state $|r_0\rangle$ with a large laser detuning $\Delta=\eta\Omega>0$, where $\eta\gg1$. (ii) Use strong microwave fields to transfer $|r_0\rangle$ to another Rydberg state $|r_1\rangle$, where the ratio of Rydberg interaction coefficients $\varkappa=C_6(r_1r_1)/C_6(r_0r_0)<0$, i.e., all pairwise Rydberg interaction $V$ reverse their signs. (iii) For another duration $t_0/|\varkappa| $, dress the state $|r_1\rangle$ with laser detuning $\varkappa \Delta$ and Rabi frequency $\varkappa \Omega$. When decoherence is negligible, time-reversed dynamics can be observed because the two many-body states at the times of $\overline{t}_1$ and $t_0+t_{\mu}+\overline{t}_1/|\varkappa| $ coincide with each other for any $\overline{t}_1<t_0$, where $t_{\mu}$ is the duration of step (ii); A numerical simulation of this dynamics is provided in Appendix~\ref{appA}. If both $|r_0\rangle$ and $|r_1\rangle$ are s or d-orbital states, step (ii) can be realized as $|r_0\rangle\xrightarrow[\Omega_{\mu1}]{}|p\rangle\xrightarrow[-i\Omega_{\mu2}]{}|r_1\rangle$ via an intermediate Rydberg state $|p\rangle$~\cite{Anderson2014}. For brevity, we denote this complete resonant two-photon transition by $|r_0\rangle\xrightarrow[i\Omega_{\mu}]{}|r_1\rangle$ with an effective Rabi frequency $\Omega_{\mu}=[\sum_{k=1}^2\Omega_{\mu k}^2]^{1/2}/2$.

\begin{figure}
\includegraphics[width=3.4in]
{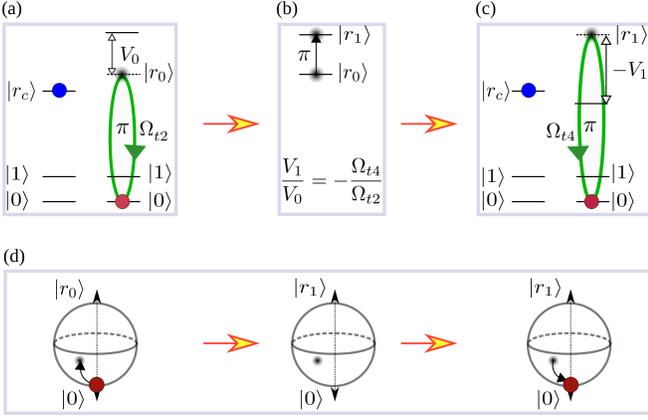}
 \caption{Refocus of a two-qubit state $|r_c0\rangle$ in a three-step time-reversal sequence. The state evolution of the target qubit is illustrated by using atomic energy levels in (a), (b), and (c), and by using the Bloch sphere in (d), respectively. (a)~[(c)] shows a $\pi$-pulse rotation between the ground state $|0\rangle$ and the Rydberg state $|r_{0(1)}\rangle$ of the target, where the dashed line highlights that the state $|r_{0(1)}\rangle$ is no longer resonant because of the positive~(negative) energy shift $V_0(-V_1)$ in the state $|r_cr_{0(1)}\rangle$. (b) shows the transition $|r_0\rangle\rightarrow |r_1\rangle$ via a large microwave Rabi frequency $i\Omega_\mu$. The laser Rabi frequencies in (a) and (b) satisfy the condition $\Omega_{t4}= -\Omega_{t2}V_1/V_0$, so that the phase accumulations for different irrational Rabi oscillations during the first and last $\pi$ pulses cancel with each other. \label{figure01} }
\end{figure}

The above theory can be used to eliminate the blockade error of the traditional Rydberg quantum gate~\cite{PhysRevLett.85.2208}, which is a major bottleneck of its achievable fidelity~\cite{Saffman2010,Saffman2016,Weiss2017}. The traditional Rydberg gate works as follows: for two quantum bits, denoted by c~(control) and t~(target) and each with basis states $\{|0\rangle, |1\rangle\}$, if one applies (I) a $\pi$ pulse to the control qubit for the rotation $|0\rangle_c\rightarrow-i|r_c\rangle_c$, (II) a $2\pi$ pulse to the target qubit for the rotation $|0\rangle_t\rightarrow-i|r_0\rangle_t\rightarrow-|0\rangle_t$, and (III) a third $\pi$ pulse similar to the first one to the control, the four independent two-qubit input states will evolve as $\{|00\rangle,|01\rangle,|10\rangle,|11\rangle \}\rightarrow- \{|00\rangle,|01\rangle,|10\rangle,-|11\rangle \}$. The key step is an excitation blockade $|r_c0\rangle\nleftrightarrow|r_cr_0\rangle$ for the input state $|00\rangle$ during the $2\pi$ pulse, where the state $|r_cr_0\rangle$ is shifted away from resonance by an interaction $V_0$~[see Fig.~\ref{figure01}(a)]. This blockade process, however, is accurate only when $V_0$ is infinitely large compared with the Rabi frequency $\Omega_{t2}$ of the $2\pi$ pulse. For any realistic (hence, finite) $V_0$, there is a population loss $\epsilon_1\sim \Omega_{t2}^2/V_0^2$ in the state $|r_c0\rangle$ resulting from even a marginal breakdown of $|r_c0\rangle\nleftrightarrow|r_cr_0\rangle$. To bring back the lost population~($\sim \epsilon_1\ll1$) to the ground state, one can replace the $2\pi$ pulse in step (II) by another set of three $\pi$ pulses upon the target qubit: [II(a)] First, apply a laser $\pi$ pulse for the rotation $|0\rangle_t\rightarrow-i|r_0\rangle_t$ with a Rabi frequency $\Omega_{t2}$, see Fig.~\ref{figure01}(a). [II(b)] Next, a microwave $\pi$ pulse is used to couple $|r_0\rangle_t$ to a nearby Rydberg state $|r_1\rangle_t$, where the effective Rabi frequency is much larger than the Rydberg blockade, and the interaction $-V_1$ of the state $|r_cr_1\rangle$ must have a sign opposite to that of $|r_cr_0\rangle$, see Fig.~\ref{figure01}(b); this step is subjected to another blockade error $\sim \epsilon_2\ll1$. [II(c)] A laser $\pi$ pulse is then applied between $|0\rangle_t$ and $|r_1\rangle_t$, but with a negative Rabi frequency $\Omega_{t4}=-\Omega_{t2}V_1/V_0$, see Fig.~\ref{figure01}(c). Following these three steps, the lost population in the state $|r_1\rangle_t$ will be brought back to $|0\rangle_t$, with an error $\sim \epsilon_1 \epsilon_2$ that is much smaller than the standard blockade error of the gate when $\epsilon_1, \epsilon_2\sim 10^{-3}$~\cite{Zhang2012,Shi2017pra}. As will be shown later, the fidelity of the process in Fig.~\ref{figure01} remarkably does not depend on the accuracy of the van der Waals interaction~(vdWI) as long as we have $\Omega_{t4}/\Omega_{t2}=-V_1/V_0$. This can easily be satisfied for atoms cooled to the $\mu$-K regime: the ratio between $V_1$ and $V_0$ is fixed to that of the C-six vdWI coefficients when the drift of the two-qubit spacing between the time-reversal pulses of Fig.~\ref{figure01}(a) and (c) is negligible.

\begin{figure}
\includegraphics[width=3.3in]
{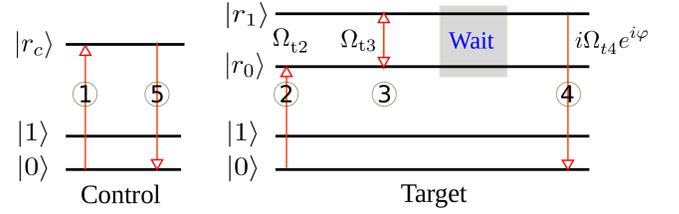}
 \caption{Sequence of the state excitation in the spin-echo controlled-phase gate. Horizontal lines denote atomic eigenstates, vertical arrows denote $\pi$-rotations between atomic states, and each number inside a circle shows the order of the $\pi$ pulse it accompanies. There is a wait duration $T=\varphi/V_1$ between the third and fourth pulses. When $\varphi=\pi$, a $C_Z$ gate is realized. The Rabi frequency in the third pulse is much larger than the energy shift $|V_{0(1)}|$ of the state $|r_cr_{0(1)} \rangle$, but those in the second and fourth are much smaller than $|V_{0(1)}|$ so as to satisfy the blockade condition. \label{figure02} }
\end{figure}

\section{Spin-echo controlled-phase gate}

    Below, we present a protocol consisting of five $\pi$ pulses and one wait that transforms the two-qubit computational basis $\{|00\rangle,|01\rangle,|10\rangle,|11\rangle \}\rightarrow\{-|00\rangle,-|01\rangle,e^{-i\varphi }|10\rangle,|11\rangle \}$, where the phase $\varphi$ is determined by the phases of the laser fields in the fourth optical pulse as shown in Fig.~\ref{figure02}, and $\varphi=\pi$ leads to a controlled-Z~($C_Z$) gate. For simplicity, we label the $k$th $\pi$ pulse by pulse-k or \circled{k} when appropriate. The Hamiltonian is $\hat{H} =\hat{H}_{\text{v}}+\hat{H}_c+ \hat{H}_t$, 
where $\hat{H}_c$ and $\hat{H}_t$ are single-atom Hamiltonians for the control and target, respectively, and $\hat{H}_{\text{v}}$ denotes the two-atom vdWI, given by
\begin{eqnarray}
\hat{H}_{\text{v}}&=&  V_0 |r_{c}r_0\rangle\langle r_{c}r_0 | -  V_1 |r_{c}r_1\rangle\langle r_{c}r_1 |,\nonumber\\
  \hat{H}_c &=& \Omega_{c}|r_c\rangle_c \langle 0| /2+\text{H.c.} ,~~~~~~~~~~~\text{pulse-1\&5},\nonumber\\
  \hat{H}_t &=&\left\{ \begin{array}{ll} \Omega_{t2} |r_0\rangle_t \langle 0| /2+\text{H.c.} ,&\text{pulse-2},\\ \Omega_{t3} |r_1\rangle_t  \langle r_0| /2 +\text{H.c.},&\text{pulse-3},
    \\ i \Omega_{t4} e^{i\varphi} |r_{1}\rangle_t  \langle 0| /2 +\text{H.c.},&\text{pulse-4},
  \end{array}\right. 
  \label{hamiltonian0}
\end{eqnarray}
where $V_0= C_6(r_cr_0)/L^6$ and $V_1= -C_6(r_cr_1)/L^6$ are positive, $\varphi=  V_{1}T$, and $T$ is a wait duration between pulse-3 and 4 that satisfies $V_1T=\pi,3\pi,5\pi,\cdots$ for a $C_Z$ gate. Here $\Omega_{t4} / \Omega_{t2} = |\varkappa|$ is required to realize a time-reversed state evolution, where $ \varkappa =C_6(r_cr_1)/C_6(r_cr_0) $. According to the Rydberg blockade mechanism~\cite{PhysRevLett.85.2208}, the Rabi frequencies during pulse-2 to 4 should satisfy
\begin{eqnarray}
\Omega_{t2}, \Omega_{t4} \ll V_0, V_1;~~V_+\equiv V_0+V_1\ll\Omega_{t3}. \label{condition}
\end{eqnarray}

Among the four input states, $|11\rangle$ remains unchanged, and $|01\rangle$ and $|10\rangle$ are optically excited but not influenced by Rydberg blockade. Thus, one can verify that in the ideal case the gate transforms them according to
\begin{eqnarray}
  && |01\rangle  \xrightarrow  [ ]{\circled{1}}  -i  |r_c1\rangle  \xrightarrow  [ ]{\circled{5}} -  |01\rangle,\nonumber\\
  && |10\rangle  \xrightarrow  [ ]{\circled{2}}  -i  |1r_0\rangle  \xrightarrow  [ ]{\circled{3}}  -  |1r_1\rangle  \xrightarrow  [ ]{\circled{4}} e^{-i\varphi}  |10\rangle. \label{state10and11}
\end{eqnarray}

Only the input state $|00\rangle$ experiences Rydberg blockade from pulse-2 to pulse-4. After pulse-1, $|00\rangle$ is mapped to $-i|r_c0\rangle$. The Hamiltonian during pulse-2~[see Eq.~(\ref{hamiltonian0})] can be diagonalized as $H_{00}^{(\text{p2})}= \sum_{\chi=\pm}\epsilon_\chi |v_\chi \rangle\langle v_\chi |$ with the following eigenvalues and the eigenvectors~\cite{Shi2017}
\begin{eqnarray}
  \epsilon_\pm &=& (V_0\pm \bar\Omega_{t2} )/2,\nonumber\\
  |v_\pm \rangle &= & (\frac{\Omega_{t2} }{2} |r_c0\rangle + \epsilon_\pm |r_cr_0\rangle)/N_\pm,\label{eigensystem01}
\end{eqnarray}
where $N_\pm  = \sqrt{[\Omega_{t2} ]^2/4 + \epsilon_\pm^2}$ and $\bar\Omega_{t2} \equiv\sqrt{[\Omega_{t2} ]^2+V_0^2}$. Below, we use $t_k$ to denote the duration of pulse-k. Then, the state $|\psi(t_1)\rangle=-i|r_c0\rangle$ at the beginning of pulse-2 can be written as $i(\sin\alpha|v_+ \rangle+\cos\alpha|v_-\rangle )$, and its state evolution leads to~\cite{Shi2017}
\begin{eqnarray}
  |\psi(t_1+t_2)\rangle &=&ie^{-i\frac{\pi\epsilon_+}{ \Omega_{t2} }} \sin\alpha|v_+ \rangle + i e^{-i\frac{\pi\epsilon_-}{ \Omega_{t2} }}\cos\alpha |v_- \rangle,\nonumber\\
  \label{statet1t2}
\end{eqnarray}
at the end of pulse-2. Because the two phase terms $-\frac{\pi\epsilon_\pm}{ \Omega_{t2}}$ accompanying the components $|v_\pm \rangle$ in $|\psi(t_1+t_2)\rangle$ cannot simultaneously be multiples of $2\pi$ due to the irrational term $\bar\Omega_{t2} $ in $\epsilon_\pm$, a small, undesirable phase factor arises in the ground-state component in $|\psi(t_1+t_2)\rangle$ along with a population leakage, $\epsilon_1=\kappa_2^2\sin^2(\pi/2\kappa_2)$, into the state $ |r_cr_0\rangle$, where $\kappa_2=\Omega_{t2} /\bar\Omega_{t2} $.

Such a population loss in pulse-2 can be restored with the subsequent pulses-3 and 4, with the phase error in the ground state eliminated simultaneously. The argument is as follows: When $\Omega_{t3}\gg V_+$, pulse-3 transfers the small population of the target in the Rydberg state from $|r_0\rangle$ to $|r_1\rangle$, so that the component $|r_cr_0\rangle$ in $|\psi\rangle$ becomes $-i|r_cr_1\rangle$. The probability for pulse-3 to fail is $\epsilon_2= 1-\kappa_3^2\sin^2(\pi/2\kappa_3)\ll1$, where $\kappa_3=\Omega_{t3}/\bar\Omega_{t3}$ and $\bar\Omega_{t3}\equiv\sqrt{\Omega_{t3}^2+V_+^2}$. Following this, a wait duration $T$ is allowed to elapse so that the two eigenstates $|v_\pm \rangle$ in $|\psi\rangle$ evolve to
\begin{eqnarray}
|\tilde{v}_\pm \rangle \equiv(\frac{\Omega_{t2} }{2} |r_c0\rangle -ie^{iV_1T} \epsilon_\pm |r_cr_1\rangle)/N_\pm,
 \label{v02}
\end{eqnarray}
which coincide with the eigenstates for the following Hamiltonian in the subsequent pulse-4 when $V_1T=\varphi$,
\begin{eqnarray}
  H_{00}^{(\text{p4})}&=& 
    (ie^{i\varphi}\Omega_{t4}|r_cr_1\rangle\langle r_c0|+\text{H.c.}  ) /2 -V_1|r_cr_1\rangle \langle r_cr_1| .
\nonumber\\     \label{pulse5}
\end{eqnarray}
More important, the eigenvalues of Eq.~(\ref{pulse5}) are $-\Omega_{t4}\epsilon_\pm/\Omega_{t2} $ for the states $|\tilde{v}_\pm \rangle$, which means that the two phases $-\frac{\pi\epsilon_\pm}{ \Omega_{t2}}$ in Eq.~(\ref{statet1t2}) can be exactly compensated during pulse-4, leading to
\begin{eqnarray}
  |\psi(\sum_{j=1}^4t_j+T)\rangle &=&i \sin\alpha|\tilde{v}_+ \rangle +i \cos\alpha |\tilde{v}_- \rangle=-i|r_c0\rangle.
\end{eqnarray}

After pulse-5 has pumped the control qubit back to the ground state, the state evolution $|00\rangle\rightarrow-|00\rangle$ is complete. To summarize, pulses 1-5 induce the following transformation in the computational basis $\{|00\rangle,|01\rangle,|10\rangle,|11\rangle \}$
\begin{eqnarray}
U_{C_\varphi}&=& \text{diag}\{ -1, -1, e^{-i\varphi },1\},\label{Cnotvariant}
\end{eqnarray}
which, after shifting the phase of the control qubit state $|0\rangle$ by $\pi$, becomes a controlled phase gate. It changes the phase of the target qubit state $|0\rangle$ by $-\varphi$ only if the control qubit is initialized in $|1\rangle$, and it becomes a $C_Z$ gate when $\varphi=\pi$. Notice that $\varphi$ is determined not by the actual Rydberg blockade, but instead by the phases of laser fields, which is set according to the qubit spacing and the C-six coefficients of the vdWI. If a gap time $T_{\text{gap}}$ is inserted between pulse-2 and pulse-3, a phase correction $-V_0T_{\text{gap}}$ should be made to the Rabi frequency of pulse-4. Furthermore, if external fields introduce Rabi frequencies $\Omega_{t2}e^{i\varphi_2 }$ and $\Omega_{t3}e^{i\varphi_3 }$ in pulse-2 and pulse-3, respectively, the Rabi frequency during pulse-4 becomes $i \Omega_{t4} e^{i\varphi'}$ with $\varphi'=\varphi +\varphi_2+\varphi_3-V_0T_{\text{gap}} $. This means that only the phase of the laser fields used in pulse-4 should be tuned. In this case, a $C_Z$ gate is realized when $\varphi'/\pi$ is an odd integer.

The above gate protocol is not based on a symmetric interaction like $-V_1/V_0=-1$, but is applicable as long as $-V_0V_1<0$. This can be understood from the spin-echo nature of the gate protocol. In the picture of the two-level system in Eq.~(\ref{H00}), spin echo is achieved when the time-evolution operators for the forward and backward nutation, i.e., $U_{\circlearrowright}=$exp$(-i\hat{H}t_0)$ and $U_{\circlearrowleft}=$exp$(-i\hat{H}'t')$, are inverse to each other, which is achievable whenever $\hat{H}'t'+\hat{H}t_0=0$. Figure~\ref{figure01} highlights the core of our gate protocol, where the spin echo is realized by setting $\Omega_{t4}/\Omega_{t2}=-V_1/V_0$, so that a $\pi$ pulse for the forward rotation and a $\pi$ pulse for the backward rotation satisfy the `echo' condition of $\hat{H}'t'+\hat{H}t_0=0$.

\begin{figure}
\includegraphics[width=3.3in]
{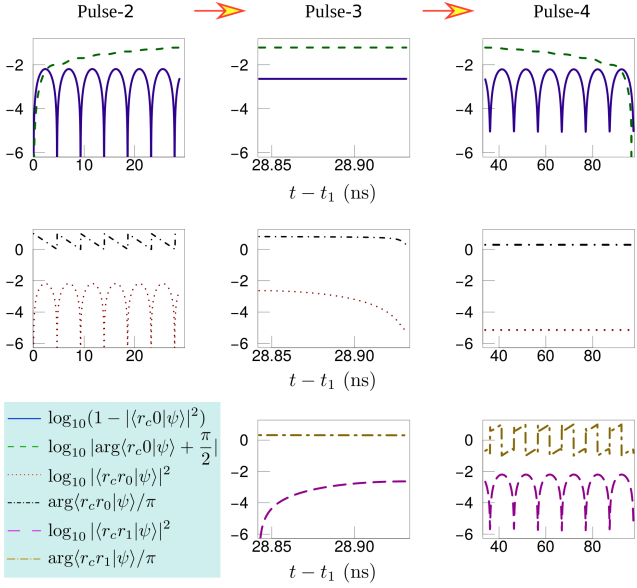}
 \caption{Population and phase dynamics of the three components $|r_c0\rangle,|r_cr_0\rangle$, and $|r_cr_1\rangle$ in the wavefunction $|\psi\rangle$ during pulses 2-4 of the gate sequence when the input state is $|00\rangle$. Calculation is performed by using $\hat{H}_{\text{v}}$ and $\hat{H}_t$ in Eq.~(\ref{hamiltonian0}) on the the initial state $-i|r_c0\rangle$, with parameters (see Sec.~\ref{sec03}) $[C_6(r_cr_0),~C_6(r_cr_1)]/2\pi=[56.2,-25.6]$~THz$\mu m^6$, $L=8\mu$m, $(\Omega_{t2},\Omega_{t3},\Omega_{t4} )=V_+( 1/\eta,\eta,|\varkappa|/\eta)$, and $\eta=18$. Here arg$(\cdot)$ gives the argument of a complex variable. Because the population involved in the transfer and the deviation of the argument of $\langle r_c0|\psi\rangle$ from $-\pi/2$ are tiny, we use the common logarithm to show the population transfer and phase dynamics for the state coefficient of $|r_c0\rangle$. Similarly, the populations in $|r_cr_0\rangle$ and $|r_cr_1\rangle$ are also small and shown with the common logarithm. The periodic population and phase evolution during pulses 2(4) has a period of $2\pi/\bar\Omega_{t2(t4)}$ because $|\epsilon_+-\epsilon_-|=\bar\Omega_{t2(t4)}$~[see Eq.~(\ref{eigensystem01})]. The values of arg$\langle r_cr_1|\psi\rangle$ at the end of pulse-3 and at the beginning of pulse-4 differ by $V_1T$~[see Eq.~(\ref{v02})].     \label{figure-3state} }
\end{figure}

\begin{figure}
\includegraphics[width=3.3in]
{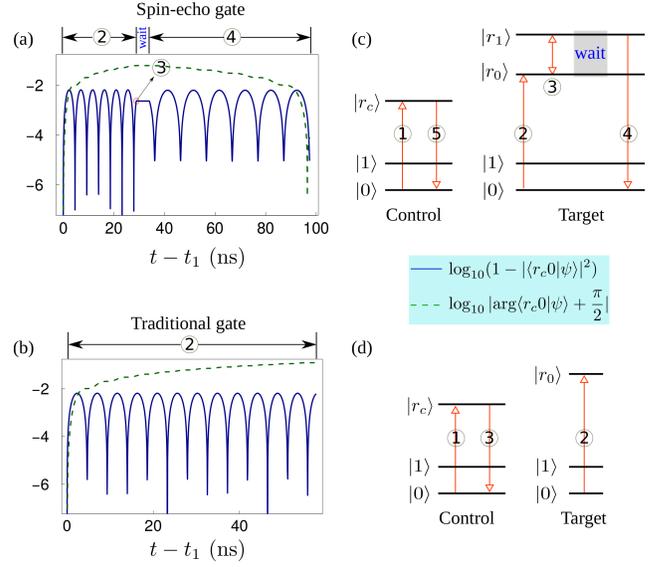}
\caption{Population and phase dynamics of the state component $|r_c0\rangle$ in the wavefunction from pulse-2 to pulse-4 of the spin-echo gate (a), and during the second pulse of the traditional gate (b). (c) and (d) show the gate sequences for the spin-echo gate and the traditional gate, respectively. The Rabi frequency for the second pulse in (b) [or (d)] is equal to that for pulse-2 in (a) [or (c)]. Other parameters are the same with those in Fig.~\ref{figure-3state}. Compared to (a), the population and phase error at the right edge of (b) are much larger. Note that there is a duration of pulse-3 in (a) that is short compared to those of pulses 2 and 4. \label{figure-comparison} }
\end{figure}

\begin{figure}
\includegraphics[width=3.3in]
{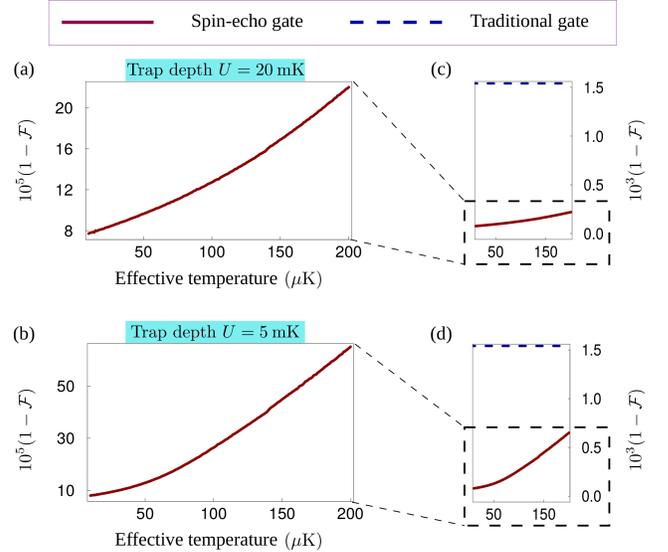}
 \caption{(a) [(b)] shows the error of the spin-echo $C_z$ gate scaled up by $10^5$ when the trap depth $U$ is $20[5]$~mK. For comparison, the solid and dashed curves in (c,d) show the gate errors scaled up by $10^3$ for the spin-echo protocol and the traditional protocol, respectively. Parameters used in the numerical calculation are the same with those in Figs.~\ref{figure-3state} and~\ref{figure-comparison}, except that here we assume a thermal distribution of the atomic location in the dipole trap and consider drift of the atoms during the gate sequence.    \label{figure03} }
\end{figure}

\section{Gate error}\label{sec03}
In this section, we systematically investigate the error of a spin-echo $C_Z$ gate~[$\varphi=\pi$ in Eq.~(\ref{Cnotvariant})] by considering, for example, a system of two $^{87}$Rb atoms. We take the qubit states $|0(1)\rangle= |5s_{1/2},F=1(2),m_F=0\rangle $~\cite{Isenhower2010} and the Rydberg states $|r_c\rangle=|r_0\rangle$ and $|r_{0(1)}\rangle =|100(105)s_{1/2},m_J=1/2,m_I=3/2\rangle$, where the interaction coefficients are $[C_6(r_cr_0),~C_6(r_cr_1)]/2\pi=[56.2,-25.6]$~THz$\mu m^6$~\cite{Shi2014}. For the external control fields, we choose $\Omega_c/2\pi=10$~MHz and $(\Omega_{t2},\Omega_{t3},\Omega_{t4} )=V_+( 1/\eta,\eta,|\varkappa|/\eta)$ with a large $\eta=18$ to fulfill Eq.~(\ref{condition}). To show that the fluctuation of the vdWI caused by qubit drift is not a severe issue, we consider a relatively small two-qubit spacing of $L=8\mu$m. For comparison, we also study the gate error of a traditional $C_Z$ gate realized by replacing pulse-2 to pulse-4 in our spin-echo gate with a $2\pi$ pulse of duration $2\pi/\Omega_{t2}$.

Below, we first study how the spin echo suppresses the blockade error in Sec.~\ref{sec03A}, and then look at the numerically calculated gate error caused by the imperfect blockade, the Rydberg state decay, and the variation of the Rydberg interaction in Sec.~\ref{sec03B}. We then discuss the gate error caused by motion induced dephasing in Sec.~\ref{sec03C}.

\subsection{Suppression of the blockade error}\label{sec03A}
The main advantage of the spin echo gate sequence is its ability to suppress the blockade error, which is achieved by pulses 2-4. However, there is a residual blockade during pulse-3 so that not all population in $|r_cr_0\rangle$ is transferred to $|r_cr_1\rangle$, leading to a small rotation error for the gate protocol. To examine this issue under realistic experimental conditions, we study, for the parameters mentioned above, the populations in $|r_c0\rangle,|r_cr_0\rangle$, and $|r_cr_1\rangle$ by using base-10 logarithm during pulses 2-4 in Fig.~\ref{figure-3state}. As discussed previously, the lost population~$(\approx\epsilon)$ in $|r_c0\rangle$ at the end of pulse-2 can be restored by the spin echo sequence of pulse-3, the wait, and pulse-4, with an error of about $\epsilon^2$. As shown in the logarithms in Fig.~\ref{figure-3state}, the population error
in $-i|r_c0\rangle$ at the end of pulse-2, of order of $10^{-3}$, is indeed reduced to the order of $10^{-6}$ at the end of pulse-4. In principle, this rotation error can be further suppressed with a larger Rabi frequency $\Omega_{t3}$ by using strong enough microwave fields. Figure~\ref{figure-3state} also shows that the deviation of the phase of $\langle r_c0|\psi\rangle$ from $-\pi/2$ is substantially suppressed by the spin echo sequence.

A comparison between the traditional gate and our spin-echo gate readily shows why the latter protocol is much more effective at suppressing the blockade error. In Fig.~\ref{figure-comparison}(a), we show the population and phase errors of the state component $|r_c0\rangle$ in the wavefunction during pulses 2, 3, the wait period, and pulse-4 of the spin-echo gate. For comparison, Fig.~\ref{figure-comparison}(b) shows the time evolution of $\langle r_c0|\psi\rangle$ during the $2\pi$ pulse of the traditional gate. From the schematics of  the gate sequences for the spin-echo gate and the traditional gate in Figs.~\ref{figure-comparison}(c) and~\ref{figure-comparison}(d), respectively, one can understand that, for both protocols, the blockade error occurs only when the target qubit is pumped. This is shown in Figs.~\ref{figure-comparison}(a) and~\ref{figure-comparison}(b) for the spin-echo and traditional cases, respectively. Here, we emphasize again that logarithmic scales are used in Fig.~\ref{figure-comparison}(a) and~\ref{figure-comparison}(b). In particular, one finds that the population and phase errors of the wavefunction in the traditional gate are orders of magnitude larger than those in the spin echo gate when excitation of the target qubit completes. Although the phase error can be removed by adjustment of the phases of the laser fields~\cite{Zhang2012}, the population loss inevitably leads to errors on the order of $10^{-3}$.

\subsection{Rydberg state decay and fluctuation of vdWI}\label{sec03B}
Firstly, decay of Rydberg state induces an error of about $E_{\text{de}}=\sum_{\alpha}T_{\text{Ry}\alpha}/\tau_\alpha$, where $T_{\text{Ry}\alpha}$ is the average time for the qubit-system to be in the Rydberg state $|\alpha\rangle$ of lifetime $\tau_\alpha$. Secondly, errors can arise if $V_1T$ in Eq.~(\ref{v02}) deviates from $\varphi$ in Eq.~(\ref{pulse5}) when the qubit spacing fluctuates. This can be studied numerically by assuming a thermal distribution of the qubit position in an optical tweezer~\cite{foot02} where the qubit spacing $L'$ during pulse-2 can deviate from the value of $L$. Thirdly, when optical traps for atoms are turned off during the gate sequence, the gate fidelity can also be hampered by the change of qubit spacing from pulse-2 to pulse-4. For an approximate analysis for the worst cases denoted by $\beta=1(-1)$ when the two qubits depart from~(approach) each other along the quantization axis, the average qubit spacing $L'$ for pulse-2 becomes $L'+ 2\beta v_z t_T$ during the wait period of the sequence, which further becomes $L'+2\beta v_z  t_{p4}$ during pulse-4, where $v_z = \sqrt{k_BT_a/m_a}$ is the r.m.s. speed of the atom along the quantization axis~($\overline{v_z^2}=\overline{v^2}/3$), and $k_B,~T_a$, and $m_a$ are the Boltzmann constant, atomic temperature, and the mass of a qubit, respectively. Here $t_T=\pi[1/2\Omega_{t2}+1/\Omega_{t3}+T/2\pi]$ and $t_{p4}=t_T+\pi/2\Omega_{t4}+T/2$.

We use the Hamiltonian in Eq.~(\ref{hamiltonian0}) to simulate the rotation error defined as $E_{\text{ro}} = \sum_{\beta=\pm1}(1-|\langle 00|U_{CZ}^\dag e^{-i\int\hat{H}t} |00\rangle|^2)/8$. The total gate error $E_{\text{de}}+E_{\text{ro}}$ as a function of $T_a$ is shown in Fig.~\ref{figure03}(a) and~\ref{figure03}(b) for traps with two different depths $U=20$ and $5$~mK, respectively~\cite{foot01}, where the Rydberg state decay rates are calculated by assuming an environment temperature of $4.2$~K~\cite{Beterov2009}. The dashed curve in Fig.~\ref{figure03}(c) and (d) shows the gate error~(calculated in a similar way)~for a traditional $C_Z$ gate realized by replacing pulse-2 to pulse-4 by a $2\pi$ pulse of duration $2\pi/\Omega_{t2}$. The gate error of the spin-echo $C_Z$ gate is below $10^{-4}$ when $T_a< 58\mu$K and $<31\mu$K in Fig.~\ref{figure03}(a) and Fig.~\ref{figure03}(b), respectively, demonstrating its robustness against position fluctuation and drift of the qubits. Note that we have ignored the population leakage to Rydberg levels near the states $|r_c\rangle,|r_0\rangle$, and $|r_1\rangle$. As analyzed in Ref.~\cite{Shi2017}, this loss can lead to an extra error, on the order of $10^{-5}$, to the gate fidelity. However, such an error can be effectively removed using the techniques of pulse shaping~\cite{Theis2016}, or avoided by shifting away the nearby Rydberg states through external fields~\cite{Petrosyan2017}. Thus, it is not a fundamental issue. On the other hand, the decay error for the above spin-echo gate, $E_{\text{de}}\approx7.1\times10^{-5}$, dominates the total gate error if the position fluctuation and drift of the qubits are negligible~[which corresponds to the left edges of Fig.~\ref{figure03}(a) and Fig.~\ref{figure03}(b)]. This observation leads to the conclusion that the Rydberg state decay sets a fundamental limit to the achievable gate accuracy for our spin-echo Rydberg quantum gates~\cite{Saffman2016}.

\subsection{Motion-induced dephasing}\label{sec03C} 
As reported in Ref.~\cite{Wilk2010}, the thermal distribution of the atomic qubit in an optical dipole trap can effectively induce a randomly distributed relative phase between the two $\pi$ pulses upon the control qubits, resulting in a phase error in the two-qubit states when the gate sequence completes. This can be called Doppler dephasing. Compared to other noises such as fluctuations of the magnetic field, Doppler effect is the dominant dephasing problem and can substantially reduce the fidelity of a quantum control involving gap times between excitation and de-excitation of Rydberg states~\cite{Wilk2010,Saffman2011,Saffman2016}. If the mass of the qubit used is $m$, the wave-vector used for the excitation of Rydberg states is $k$, Ref.~\cite{Saffman2011} estimated that there is an error of $E_{\text{Do}}\approx[1-e^{-k_BT_a(kt)^2/2m}]/2$ for a state evolution during which a Rydberg atom drifts for a time of $t$. Because smaller $k$ reduces this dephasing effect, the excitation of a high-lying Rydberg state of even parity by a two-photon process is more favorable compared to using a $p$-orbital state through an ultraviolet single-photon process. For a configuration of two counterpropagating optical laser fields for exciting $s$ or $d$-orbital Rydberg states, Ref.~\cite{Saffman2016} showed that cooling qubits to several tens of $\mu$K is enough to reduce $E_{\text{Do}}$ below the level of $10^{-4}$ when $t\lesssim100$~ns for a cesium-atom Rydberg gate. For the example shown in Figs.~\ref{figure-3state},~\ref{figure-comparison}, and~\ref{figure03}, the control qubit is in the Rydberg state for a time of $t\sim 97$~ns from pulse-2 to pulse-4 if the input state is $|00\rangle$ or $|01\rangle$. Since the wavevectors of Rydberg lasers are similar for rubidium and cesium, it is possible to reach a small $E_{\text{Do}}$ on the order of $10^{-4}$~(or $10^{-5}$) by cooling qubits to the order of $10\mu$K~(or $1\mu$K), which means that the dephasing of atomic transition is a more severe problem compared to the variation of Rydberg blockade.

Among the numerous sources of gate error, our study shows that the motion of qubits is the main concern for the state-of-the-art techniques in which experiments on Rydberg gates use optical dipole traps for qubits. First, it can induce fluctuation of the vdWI. However, our numerical calculation shows that the vdWI-fluctuation-induced error of a spin-echo $C_Z$ gate can be as small as several times $10^{-4}$ even if the optically trapped qubits are cooled only to the sub-mK regime. Second, the qubit motion can also dephase the atomic transition between ground and Rydberg states, and the resulting error can be reduced to the order of $10^{-4}$ only when the qubits are cooled to tens of $\mu$K for the widely adopted scheme of two-photon excitation of Rydberg states. Consequently, sufficient atomic cooling is necessary to attain the predicted fidelity of the spin-echo Rydberg gate, which is limited by decay of the Rydberg states. This latter error, in principle, can be reduced to the order of $10^{-6}$ by choosing more stable Rydberg states and stronger laser fields for faster state manipulations~\cite{Theis2016,Petrosyan2017}.

\section{Conclusions}
We study spin echo with neutral Rydberg atoms and show that it can reduce the well-known blockade error in a traditional Rydberg gate from the order of $\epsilon\sim 10^{-3}$ to $\epsilon^2$, leading to an accurate controlled-phase gate of intrinsic fidelity limited only by the Rydberg state decay. Moreover, the spin-echo gate is resilient to the fluctuation of the blockade strength between Rydberg atoms, making it possible to realize accurate two-qubit Rydberg logic gates with neither pulse shaping nor ground-state cooling of qubits.

\section*{ACKNOWLEDGMENTS}
The author thanks Yan Lu for fruitful discussions and acknowledges support from the Fundamental Research Funds for the Central Universities and the 111 Project (B17035).

\appendix{}
  \section{Time-reversed many-body spin dynamics}\label{appA}
  In this appendix, we show another application of the neutral-atom spin echo, i.e., quantum simulation of time-reversed many-body dynamics with Rydberg blockade. The numerical simulations in this appendix reveal that the `echo' of any initial state can occur as long as the spin echo sequence introduced below Eq.~(\ref{H00}) is applied. We note that an understanding of the many-body `echo' can help one to easily grasp why the blockade error of our spin-echo gate is suppressed.

In the study of quantum simulation with Rydberg blockade, a many-body Hamiltonian can be constructed either from resonant pumping~\cite{Labuhn2016} or from off resonantly dressing ground states with Rydberg states~\cite{Zeiher2016,Zeiher2017}. As an example, we consider $N$ atoms trapped in a one-dimensional optical lattice of lattice constant $\mathscr{L}$, where each atom is prepared in a superposition of two ground states $|0(1)\rangle$, and subsequently dressed with Rydberg state $|r_0\rangle$ via a detuned transition $|1\rangle\leftrightarrow|r_0\rangle$ for each atom during the time $t\in[0,~t_0]$ characterized by the following Hamiltonian,
\begin{eqnarray}
  \hat{H}&=& \sum_{k=1}^{N}\left[ \Omega (|r_0\rangle_k\langle 1|  + \text{H.c.})/2 + \Delta |r_0\rangle_k\langle r_0| \right]  \nonumber\\
  &&+\sum_{j=1}^{N-1} \sum_{k=j+1}^{N} C_6(r_0r_0)/L_{jk}^6|r_0r_0\rangle_{jk}\langle r_0r_0| ,\label{dressing}
\end{eqnarray}
where $L_{jk}$ is the distance between atoms at sites $j$ and $k$. Afterwards, a strong microwave field is applied to cause an almost complete transition $|r_0\rangle_k\rightarrow|r_1\rangle_k $ for every atom $k$ with a Rabi frequency $i\Omega_\mu$ that is much larger than the Rydberg interaction. The choice of Rydberg states should follow the condition of $\varkappa=C_6(r_1r_1)/C_6(r_0r_0)<0$. Then, Rydberg dressing $|1\rangle\leftrightarrow|r_1\rangle$ with $(\Omega,\Delta)$ in Eq.~(\ref{dressing}) replaced by $(\Omega,\Delta)\varkappa$ induces a state evolution during $t\in(0,~t_0]/|\varkappa| +t_0+\pi/\Omega_\mu$ that is exactly the time-reversal counterpart to that during $t\in(0,~t_0]$.
  
\begin{figure}
\includegraphics[width=2.8in]
{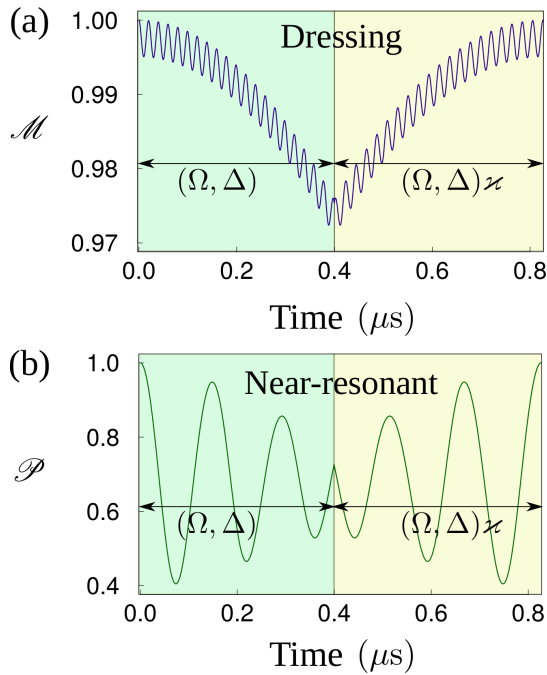}
 \caption{Time reversed dynamics of many-body observables in a four-atom system of Eq.~(\ref{dressing}). (a)~[(b)] works in the dressing~[near-resonant] regime, and shows the average population in the state $|\uparrow\rangle\equiv (|0\rangle+ |1\rangle)/\sqrt2$~[$|1\rangle$] where each atom is initialized. The forward~(backward) state evolution occurs during $(0,0.4]\mu$s and $(0.4,0.827]\mu$s, respectively, so that the many-body observables $\mathscr{M}$ and $\mathscr{P}$ also return to the initial values at the end of the spin echo sequence. We have ignored the duration of the transition $|r_0\rangle\rightarrow |r_1\rangle$ at $0.4\mu$s. See text for parameters. Note that not only $\mathscr{M}$ and $\mathscr{P}$, but any physical observable should follow a similar time reversed dynamics shown here. \label{figure04} }
\end{figure}
For a numerical test, we consider the following states of $^{87}$Rb in Eq.~(\ref{dressing}): $|0(1)\rangle = |5s_{1/2},F=1(2),m_F=1(2)\rangle$ and $|r_{0(1)}\rangle=|100s_{1/2}(d_{5/2}),m_J=1(5)/2,m_I=3/2\rangle$, where $|r_0\rangle$ can be transferred to $|r_1\rangle$ via an intermediate p-orbital state using microwave pulses, while $|1\rangle$ and $|r_{0(1)}\rangle$ can be transferred from each other by linearly~(circularly) polarized laser fields. The interaction coefficients are $C_6(r_0r_0)[C_6(r_1r_1)]/2\pi=56.2~[-52.6]$~THz$\mu m^6$. We first study the Rydberg dressing regime~\cite{Zeiher2016,Zeiher2017} by choosing $\mathscr{L}=10~\mu$m, $t_0=4\pi/\Omega$, and $10\Omega=\Delta=50\times2\pi$~MHz, and those for dressing $|r_1\rangle$ are given by $(\Omega,\Delta)\varkappa$. Starting from an initial state $|\psi(0)\rangle=\otimes_{k=1}^N|\uparrow\rangle_k$ with the spin-echo sequence described above, where $|\uparrow\rangle_k=(|0\rangle+ |1\rangle)_k/\sqrt2$, we probe the local transverse magnetization $\mathscr{M}=\sum_{k=1}^4\sum |\langle\cdots\uparrow_k  \cdots | \psi(t)\rangle|^2/4$, where the second sum is over all four-atom states $|\cdots\uparrow_k  \cdots \rangle$ when the $k$th atom is in the state $|\uparrow\rangle_k$. The result is shown in Fig.~\ref{figure04}(a), where we have ignored the duration $\pi/\Omega_\mu$ of the transition between $|r_0\rangle$ and $|r_1\rangle$ for clarity. Figure~\ref{figure04}(a) shows that the observable $\mathscr{M}$ experiences an evolution during $(0.4,0.827]\mu$s which is exactly time reversal to that during $(0,0.4]\mu$s. 

  The spin-echo induced reversible many-body dynamics can also happen outside the dressing regime. We then choose $\mathscr{L}=16~\mu$m and $\Omega/2=\Delta=2.5\times2\pi$~MHz in Eq.~(\ref{dressing}), and perform numerical simulation in a similar four-atom system as above. When every atom is initialized in $|1\rangle$, the average population in $|1\rangle$ defined by $\mathscr{P}=\sum_{k=1}^4\sum |\langle\cdots1_k  \cdots | \psi(t)\rangle|^2/4$ is plotted in Fig.~\ref{figure04}(b), where the evolution of $\mathscr{P}$ in the backward rotation is exactly time reversed to that in the forward rotation. In Fig.~\ref{figure04}, we have ignored the Rydberg state decay effect because the timescale here is much smaller than the Rydberg state lifetime, $1.2~(0.9)$~ms, of $|r_{0(1)}\rangle$ in a $4.2$~K environment~\cite{Beterov2009}.

  In Figs.~\ref{figure04}(a) and~\ref{figure04}(b), we have shown the time-reversed evolution of two observables, $\mathscr{M}$ and $\mathscr{P}$, for two sets of parameters. We do this not because only $\mathscr{M}$ and $\mathscr{P}$ can be refocused, but because it is not easy to pictorially and fully show the time-reversal dynamics of a many-body wavefunction. In fact, any observable should follow a time reversal evolution like those shown in Fig.~\ref{figure04} in the numerical examples studied above. Extending the above scenario to our gate sequence, one can understand that the removal of the blockade error from the Rydberg gate via the spin echo sequence described in the main text has used the fact that when the `echo' happens, the initial state is recovered. As a consequence, all population will return to the initial state.

%


\end{document}